\documentclass[%
reprint,
superscriptaddress,
 amsmath,amssymb,
aps,
]{revtex4-1}
\usepackage{graphicx}
\usepackage{natbib}
\usepackage{hyperref}

\begin{document}
\title[Three-Node Optical Time Synchronization]{Demonstration of Real-Time Precision Optical Time Synchronization in a True Three-Node Architecture \footnote{The use of brand names and/or any mention or listing of specific commercial products or services herein is solely for educational purposes and does not imply endorsement by the Air Force Research Laboratory or our partners, nor discrimination against similar brands, products or services not mentioned} }

\author{Kyle~W.Martin$^1$, Nader~Zaki$^2$, Nolan~Matthews$^2$, Matthew~S.~Bigelow$^1$, Benjamin~K.~Stuhl$^2$, John~D.~Elgin$^3$, and Kimberly~Frey$^{3,*}$}
\affiliation{$^1$ Blue Halo, Albuquerque, NM 87123, USA}

\affiliation{$^2$ Space Dynamics Laboratory, North Logan, UT 84341, USA}
\affiliation{$^3$ Space Vehicles Directorate, Air Force Research Laboratory, Kirtland Air Force Base, Albuquerque, NM 87117, USA}

\affiliation{$^*$ Author to whom any correspondence should be addressed.}
\email{qst@afrl.af.mil}

\begin{abstract}
Multi-node optical clock networks will enable future studies of fundamental physics and enable applications in quantum and classical communications as well as navigation and geodesy. We implement the first ever multi-node optical clock network with real-time, relative synchronization over free-space communication channels and precision on the order of 10 femtoseconds, realized as a three-node system in a hub-and-spoke topology.  In this paper we describe the system and its performance, including a first ever measurement of precision optical time synchronization between nodes with no direct communication link or causal feedback relationship. \\
\end{abstract}

\maketitle

\section{\label{sec:intro}Introduction}
The most precise optical atomic clocks have now demonstrated frequency stability at levels that reach below $1\times10^{-17}$ in only 1000~s \cite{ludlow2015,Bothwell_2019,Hinkley2013,Cambell2017,Oelker2019,Schioppo2017}. Networks of these optical clocks could be utilized in studies of fundamental physics of relativity \cite{wineland2010, Takamoto2020}, searches for ultralight \cite{Tsai2023} or topological dark matter \cite{Arvanitaki2015, Derevianko2014, Wicislo2018, Kennedy2020}, gravitational wave detection \cite{Takano2016,Kolkowitz2016}, and geodesy \cite{McGrew2018,Grotti2018}. Furthermore, a system synchronized to this level of precision has many applications, such as quantum networking \cite{Burenkov2023}, high data transfer rates over optical links \cite{Okawachi23}, and global navigation satellite systems \cite{Schuldt2021}.  However, these studies of clock networks are either (1) theoretical, requiring frequency stability of $< 10^{-15}$ at one second in conjunction with optical clock comparisons \cite{Tsai2023,Arvanitaki2015, Derevianko2014,Kolkowitz2016}, (2) rely on fiber links connecting measurement locations \cite{Takamoto2020,Kennedy2020,McGrew2018}, or require extra infrastructure in addition to fiber links such as precise optical clocks at several locations to calibrate measurement \cite{Grotti2018}.  Moreover, some of the studies of fundamental physics require co-located clocks whose precision and accuracy needs to be $< 10^{-18}$ to measure small environmental variations or leverage large environmental changes by using RF enabled time and frequency transfer to compare clocks at remote locations \cite{Wicislo2018}.  Establishing an optical clock network and leveraging high-performance free space optical synchronization systems would have great value in distributing time from an ultra-stable but bulky and power-hungry lattice or ion clock to a clock with lower size, weight, and power (SWaP) \cite{Hugo2016}, to realize increased time and frequency stability in environments where low instability clocks could not otherwise operate. Optical time synchronization could realize the following possibilities: (1) enable theory based experiments, (2) reduce clock precision burden by allowing for measurement sites to be further away with more stability than RF enabled time and frequency transfer, and (3) remove infrastructure requirements.

There have been several demonstrations of time synchronization at the femtosecond level (i.e., time deviation (TDEV) in the single-digit femtosecond range from 1~s to 1~hour) over continuous links between two-nodes \cite{Sinclair2019, Hugo2016, Deschenes2016, Isaac2018, Bigelow2019}.  Heretofore, these methods have only been demonstrated for a two-node point-to-point link, while many applications for an optical clock network inherently require multiple nodes.  Bodine \textit{et al.} \cite{Bodine2020} did make initial progress beyond a two-node, single-link system by constructing a mid-link relay. However, the relay was not itself synchronized or syntonized to the primary node, and thus diverged from the other two nodes.  Moreover, the two endpoints shared a common clock and so were only able to measure link noise rather than full synchronization.  Additionally, it is worth noting that in two-node systems verification methods are intrinsically limited to a comparison between a leader (i.e. primary) node and a follower (i.e. secondary) node; a more comprehensive test of the level of synchronization across nodes is to compare performance between two \textit{independent} follower nodes.  In what follows, we present results for the first demonstration of a true three-node, three-clock, free-space, real-time optical time synchronization network with femtosecond (fs) level performance.  Furthermore, we demonstrate, for the first time, measurement of time synchronization performance between two \textit{independent} follower nodes which have no feedback between them.

\section{\label{sec:twonode}Two-Way Optical Time Synchronization}

\begin{figure*}[ht]
\includegraphics[width=\textwidth]{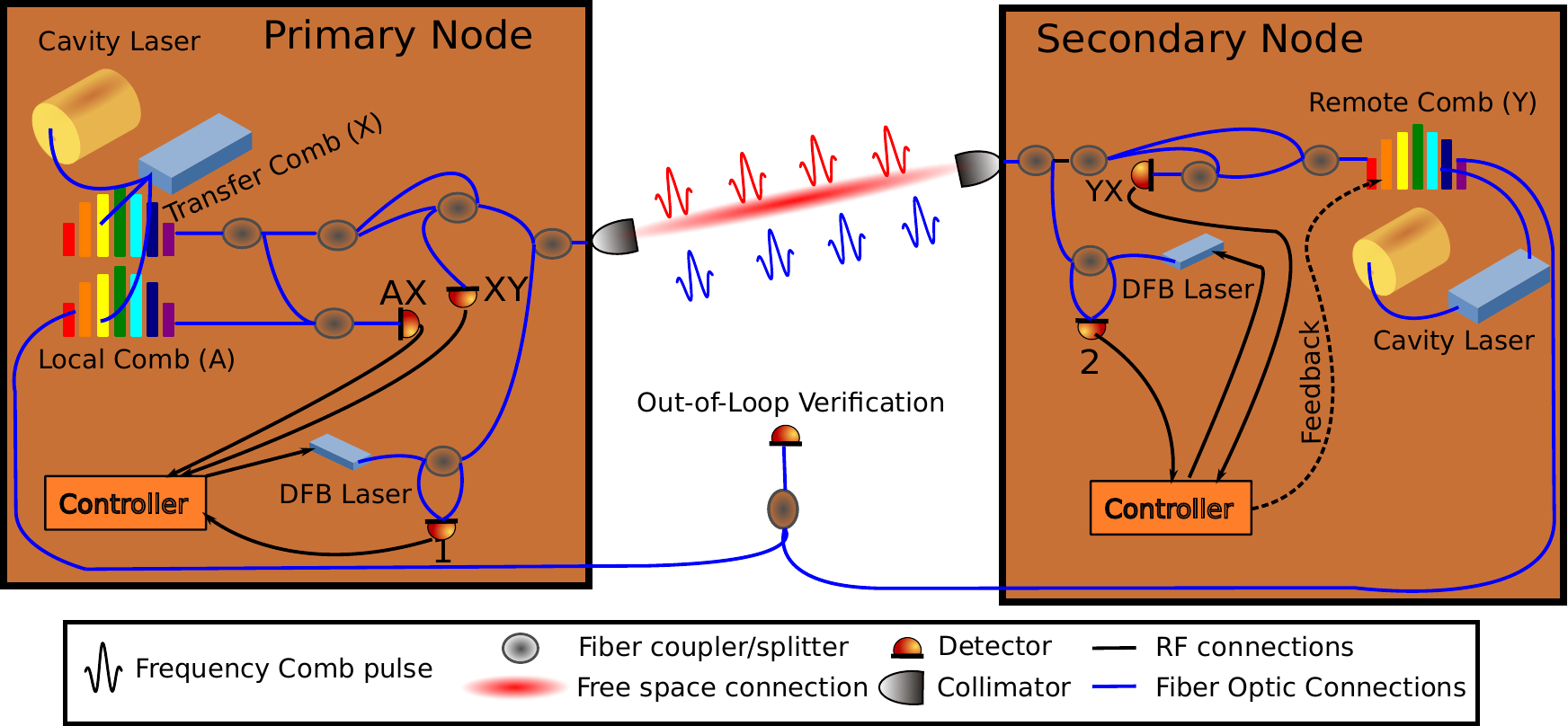}
\caption{\label{fig:twonode} Optical frequency comb-based two-way time synchronization setup for a two-node comparison.  Interferograms (IGMs) are detected at detectors XY, YX, and AX, the coarse timing information is recorded at detectors 1 and 2.   Each node contains a cavity stabilized laser and frequency comb oscillator, with the primary site containing an extra frequency comb required for cross correlation measurements.  The custom FPGA controller is responsible for calculating and communicating clock error and closing the synchronization loop at the secondary node.  For this work, we used Stable Laser Systems "Complete System" SLS-INT-1550-200-1 laser cavities operating at 1556 nm and Vescent FFC-100-200 fiber frequency combs with optical bandwidth of $\sim$100 nm centered at 1550 nm.} 
\end{figure*}

Before describing our three-node time synchronization system and results, we begin with a description of a two-node system shown in Figure~\ref{fig:twonode}, as the three-node system described in this paper is instantiated in a hub-and-spoke topology, which can be decomposed into a pair of two-node systems sharing a common primary node.  Our method of synchronization employs a two-way optical frequency comb-based time synchronization protocol similar to \cite{Sinclair2019, Hugo2016, Deschenes2016, Isaac2018, Bigelow2019} using ``linear optical sampling'' or cross-correlation measurements from optical heterodyne beats between frequency comb pulses \cite{Coddington:09}. In this scheme, local comb A at the primary site and remote comb Y at the secondary site are each individually locked to a stable reference cavity laser  such that the repetition rate of the combs is $f_{rep} \approx 200$~MHz.  These two clock combs are the physical oscillators that are synchronized to the femtosecond level.  Offset-repetition rate linear optical sampling requires an additional frequency comb, referred to as the transfer comb, which is located at the primary site and is locked with a slightly offset repetition rate $f_{rep} + \Delta f_{rep}$ where $\Delta f_{rep} \approx 2.08$~kHz.

Comb pulses are exchanged between the transfer comb X at the primary site and the remote comb Y at the secondary site over a free-space optical link. Since there is a frequency offset between the clock combs relative to transfer comb, the pulse trains will sweep through one another and produce interferograms (IGMs) at a rate of $1/\Delta f_{rep}$. In effect, this cross correlation method magnifies the time domain by a factor of $f_{rep} / \Delta f_{rep},$ enabling femtosecond resolution. The IGMs arrive at detectors XY and YX with time offsets measured at each site $\Delta\tau_{XY}$ and $\Delta\tau_{YX}$, respectively.  IGMs are generated every $1/\Delta f_{rep} \approx 481$~$\mu$s.  To properly compare the phase of the local comb A to the remote comb Y, we measure the phase offset of the local comb A with respect to transfer comb X, generating an IGM on detector AX at time $\Delta\tau_{AX}$. The clock error, $\Delta T_{A,Y}$ is described by

\begin{equation} 
\label{Eq:Master}
\Delta T_{A,Y}=\frac{1}{2} ( \Delta\tau_{YX}-\Delta\tau_{XY}) -\Delta\tau_{AX}+\tau_{plane},
\end{equation}
where $\tau_{plane}$ represents a relative offset between the clock combs outputs that is inserted in order to temporally coincide the pulses in the out-of-loop verification reference plane.  There are additional correction terms (originally derived by Desch\^{e}nes et al. \cite{Deschenes2016}) that for a static system remain constant and are accounted for in calibration.  Finally, since the repetition rate of the frequency combs leads to a timing ambiguity modulo $1/f_{rep}$,  we utilize a coarse timing link to overcome this ambiguity.  We send coarse timing information over the same free-space optical path that is used to transmit the frequency comb light; a waveguide electro-optic modulator is used to encode a pseudo random binary sequence on the phase of the continuous wave laser at 1531~nm (linewidth $<$~1~MHz), shown as the DFB laser in Figure~\ref{fig:twonode}.  This optical communications link is also the channel used to exchange the timestamps of the linear optical sampling measurements (mentioned above) enabling closure of the time synchronization loop.

Independent of the synchronization described above, an out-of-loop heterodyne method \cite{Deschenes2016} has been employed to ascertain the level of synchronization between the primary and secondary node.  This scheme entails offsetting the relative carrier envelope frequencies of the clock combs by 1~MHz relative to each other.  Assuming identical repetition rates, a consequence of this constraint is a 1~MHz heterodyne signal whose amplitude is proportional to the amount of overlap of the comb pulses at a fixed plane (i.e. a photodetector) and refers to $\tau_{plane}$ in Equation~\ref{Eq:Master}. Variations in the relative phase of the combs, induced by clock error fluctuations, appear as changes in the amplitude of the verification signal.

\section{\label{sec:experiment}Network Design}

\begin{figure*}[ht]
\includegraphics[width=\textwidth]{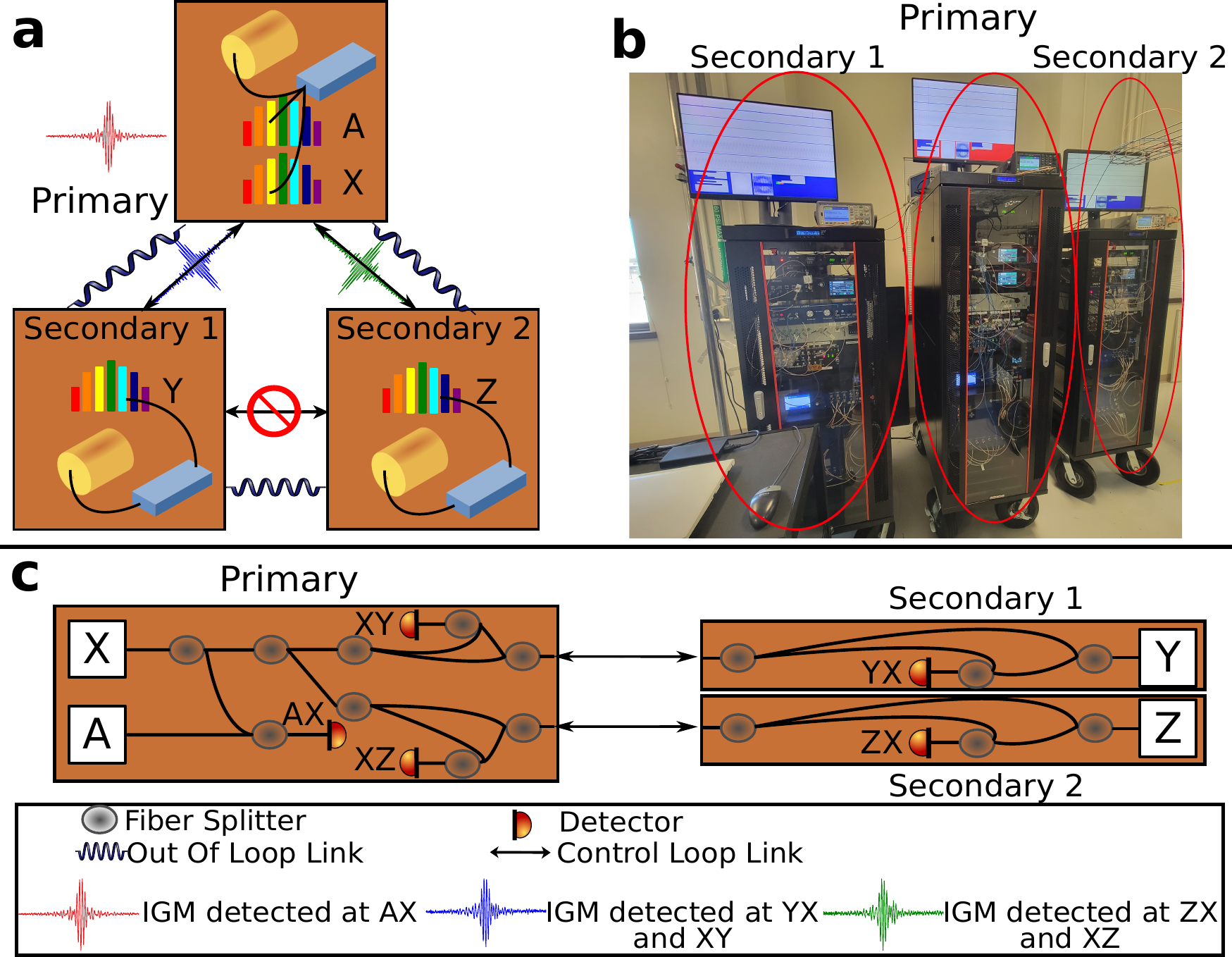}
\caption{\label{fig:hubspoke} Optical frequency comb-based time synchronization setup for three-node hub-and-spoke topology.  \textbf{a} shows the free-space optical links.  Light from the optical frequency combs at the secondary nodes are optically mixed with the synchronization comb light producing IGMs.  The time of arrival of these IGMs is measured, and the repetition rate of the optical frequency combs at each of the secondary nodes is adjusted accordingly.  These nodes are truly independent -- note there is no direct free-space link, and thus no direct control, between the secondary nodes.  Also shown here is the out-of-loop measurement performed via a fiber link in which the optical frequency combs at both secondary nodes are sent back to the primary node and compared to comb A. Furthermore, combs Y and Z are compared at both nodes 1 and 2 in a separate verification measurement.  \textbf{b} is a picture of our three-node time synchronization system.   \textbf{c} shows a simplified diagram of the IGM generation.} 
\end{figure*}

Our multi-node system, illustrated schematically in Figure~\ref{fig:hubspoke}\textbf{a}, implements a hub-and-spoke topology where the primary node distributes time to two spatially-separated secondary nodes; there is no direct optical time synchronization link between secondary sites 1 and 2.  In this experiment, all three-nodes are located in the same laboratory, but each node is self-contained within a single electronics rack and each has an independent cavity-stabilized laser (Each reference laser is a 1556~nm fiber laser locked to a cavity whose finesse is $> 3\times 10^5$, with a 3 GHz free spectral range, linewidth $<$ 1 Hz, and manufacturer estimated stability at one second of $< 3\times 10^{-15}$) as shown in Figure~\ref{fig:hubspoke}\textbf{b}.  The experiment involved two independent free-space links set up on an optics table connected to each node by 10~m of fiber.  The free-space path of the primary to each secondary node was approximately 4~m.  We also performed this experiment completely in fiber using approximately 5~m of fiber for each primary to secondary channel.  In addition to the extra equipment to set up a third node, it became necessary for an additional coarse time DFB laser at the primary node, as each pair of primary-secondary nodes required its own independent tracking of the $1/f_{rep}$ time ambiguity.  Since the hub-and-spoke node architecture is made up of two independent two-node time synchronization arms, we must duplicate Equation \ref{Eq:Master} to account for the extra secondary node and solve for the timing differences between the primary node and each secondary node.  The IGMs necessary to solve for the timing difference between the primary and secondary nodes are detected in the transceiver (Figure~\ref{fig:hubspoke}\textbf{c}) at each node.

While our implementation of a multi-node system enables the same out-of-loop measurements for characterization, it also distinctively enables characterization of the performance of the two independent secondary nodes which have no direct communication between them, as well as removes the concomitant dependency of the local clock from the characterization. In both the latter case and the standard two-node implementation, the two-nodes being compared have their $f_{ceo}$'s offset by a fixed amount (1~MHz in our case). Since we only have two independent $\Delta \tau_{plane}$ variables in our set of equations we can only guarantee that two sets of comb pulses arrive simultaneously at an arbitrary location (detector).  This motivated us to do two experiments: (1) the typical primary versus secondary comb comparison and (2) the new secondary versus secondary comb comparison.

\begin{figure*}[ht]
\includegraphics[width=\textwidth]{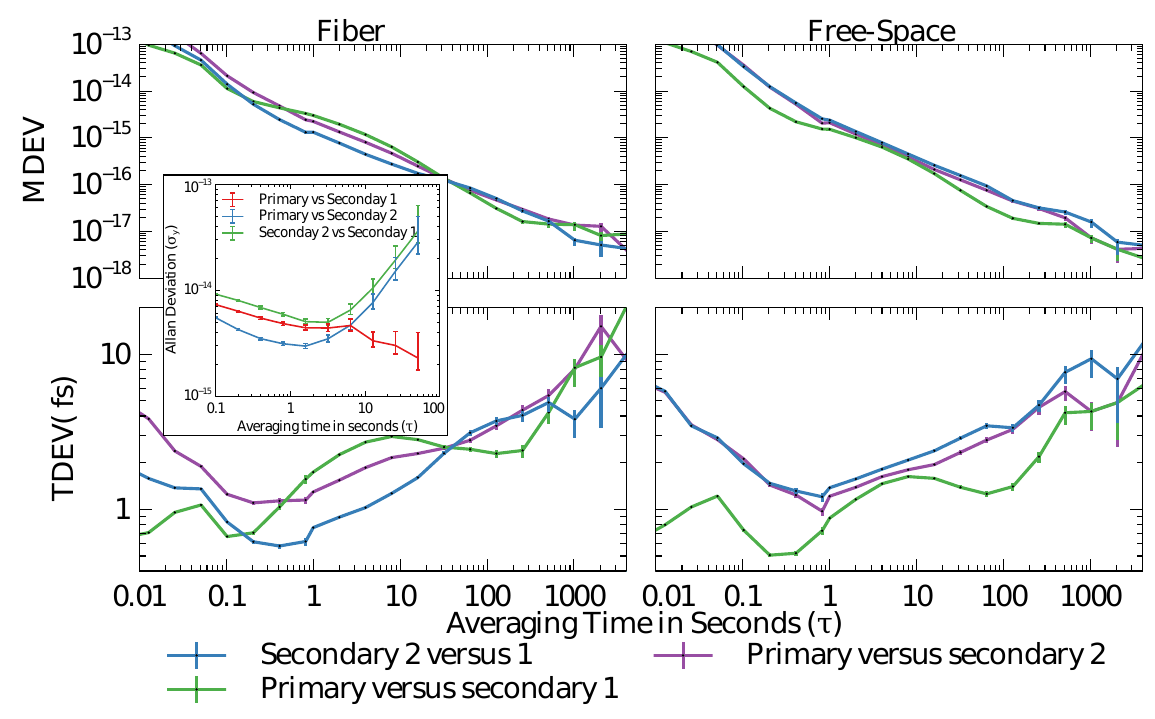}
\caption{\label{fig:results} Shown above are the results from the three-node time synchronization experiment.  We compare the primary node to each secondary node using the standard out-of-loop verification method described in the text.  We also observe the heterodyne between the two secondary combs.  Comb pulses were exchanged to synchronization time in both fiber and free-space.  Both modified Allan deviation and time deviation measurements are shown for both fiber and free space links. The right two plots show the free-space experiments which were slightly noisier than the fiber experiments (shown on the left two plots). We also note that since the system is eventually designed to operate in three distinct locations, the out-of-loop verification fiber is much longer than normal and exposed to room temperature fluctuations resulting in excess noise on the verification data. Inset in the plot is a direct heterodyne frequency stability measurement of the three unsynchronized cavity lasers. }
\end{figure*}

Figure \ref{fig:hubspoke}\textbf{c} shows a simplified optical setup of the two-way optical time synchronization experiment.  At the primary node, light from the transfer comb X is split. One part is optically mixed with local comb A and detected on detector AX generating the local/transfer IGM, while the remaining light from the transfer comb is then split again to be used separately for the two  transceivers at the primary site.  In each transceiver, the transfer comb is split one last time reserving some of the light to be optically mixed with the received remote comb light, generating the remote/transfer IGM on detectors XY and XZ, while the majority of the light is sent over the free-space optical links to each of the secondary nodes.  At each site the arrival time of the IGM is measured with respect to that site's local timebase, and the timestamps are then transmitted between sites over a free-space optical communications link co-propagating with the frequency comb pulses. The timestamps are mathematically combined to yield the time of flight between each site and the offset between each secondary site's clock and the primary site's clock. The secondary site clock combs are each adjusted in order to null the measured clock offsets. This is performed by varying the $f_{opt}$ offset frequency lock. For the secondary sites, the $f_{opt}$ signal is generated by mixing the comb-clock laser detected beat-note with a steerable tone derived from direct digital synthesis whose frequency is set by a real-time FPGA based feedback controller.

\section{\label{sec:lresults}Results}
Utilizing the out-of-loop verification technique, we characterize the performance of the time synchronization protocol with modified Allan deviation (MDEV, $\mathrm{mod}\,\sigma_{y}(\tau)$) and time deviation measurements (TDEV, $\sigma_x(\tau)$) over a few hours -- first under direct fiber connections and then over short free-space links.   The modified Allan deviation differs from the standard Allan deviation in that it numerically varies the measurement bandwidth depending on the averaging period $\tau$.  This variation of the binned time averaged phase measurements allows for the disambiguation of measurements with white phase noise and white frequency noise \cite{MDEV}.  After calculation of the modified Allan deviation the time deviation is simply given by $\sigma_x(\tau) = \frac{\tau}{\sqrt{3}}\mathrm{mod}\,\sigma_{y}(\tau)$ \cite{Handbook}.

Our results in Figure~\ref{fig:results} show TDEVs at 1 second at the femtosecond level for each of two independent links (primary $\leftrightarrow$ secondary 1 and primary $\leftrightarrow$ secondary 2), as well as similar precision for the out-of-loop comparison between the two secondary nodes.  Each data set (fiber link or free-space link) is collected under two separate initialization conditions.  Data is first simultaneously collected between the clock combs at the primary node and the two respective secondary nodes; then the offset between the two secondary node clock combs is re-adjusted to 1~MHz, and data is collected again but now between the two secondary sites.

We note that our out-of-loop measurements have excess noise relative to previously published two-node work.  As reported in \cite{sinclair2016}, the out-of-loop verification technique is susceptible to deviations in fiber path length (unlike the in-loop optical time synchronization protocol).  In this experiment the nodes were tested in the same laboratory, but each node is fully self-contained and designed to be placed at a remote location relative to the others. This design results in the out-of-loop fiber being 2~m long which has added to the residual long-term measurement noise associated with temperature fluctuations over this link.

\section{\label{sec:level2}Discussion and Future Work}
We see opportunities to greatly increase the utility of this work.  The hub-and-spoke topology used for this experiment keeps a centralized node that all other nodes are steered towards.  A more robust topology would allow any individual node to become the primary or secondary node.  Such a configuration would require an optical link between what are now the two secondary nodes and the introduction of one additional frequency comb at each of these sites.  In addition, optical switching will likely need to be implemented to determine which comb pulses are sent for the eventual heterodyne at each site.  Further optimization and improvement on the quality and precision of time synchronization is currently under investigation for the three-node system in the hub-and-spoke topology.  Next steps include increasing the length of the free-space optical links by moving the secondary nodes away from the primary while maintaining the out-of-loop verification, as well as performing measurements in non-laboratory environments.

\acknowledgments
This work is supported by the Air Force Research Laboratory. \\
\textbf{Disclaimer} Approved for public release, distribution is unlimited.  Public Affairs release approval \#AFRL20245894. The views expressed are those of the authors and do not reflect the official policy or position of the Department of the Air Force, the Department of Defense, or the U.S. government.
\section*{Data availability statement}
Data underlying the results presented in this paper are not publicly available at this time but maybe obtained from the authors upon reasonable request.

\section*{Conflict of Interest}
The authors declare no conflicts of interest.

\bibliographystyle{unsrt}
\bibliography{Three_node_paper_rewrite_revised_2}

\end{document}